\documentclass[10pt]{iopart}
\usepackage{iopams}
\usepackage{cite}
\usepackage{graphicx}
\usepackage{color}
\usepackage{siunitx}
\DeclareSIUnit\muB{$\mu_B$}
\DeclareSIUnit\atper{{at.\,\percent}}
\usepackage[normalem]{ulem}
\usepackage[colorlinks=true]{hyperref}
\usepackage{url}
\usepackage{amsfonts,amssymb}
\usepackage{nicefrac, xspace}
\newcommand{\half}{\nicefrac{1}{2}}
\newcommand{\quarter}{\nicefrac{1}{4}}
\newcommand{\tquarter}{\nicefrac{3}{4}}
\providecommand{\add}[1]{\textcolor{black}{#1}}
\newcommand{\agpd}{Ag$_c$Pd$_{1-c}$\xspace}
\newcommand{\agpdc}[2]{Ag$_{#1}$Pd$_{#2}$\xspace}
\newcommand{\Nsub}{N_\text{sub}}
\newcommand{\fwhm}{\text{FWHM}}
\newcommand{\mathperiod}{\,.} 		
\newcommand{\mathcomma}{\,,}  		
\renewcommand{\vec}[1]{{{\bi #1}}}
\newcommand{\occ}[2]{\sigma_{#1}^{#2}}
\newcommand{\op}[1]{\mathcal{#1}}

\begin{document}

\title{Quantitative description of short-range order and its influence on the electronic structure in Ag-Pd alloys}
\author{M Hoffmann$^{1,2}$, A Marmodoro$^{3}$, A Ernst$^{3}$, W Hergert$^{2}$, J Dahl$^{4,5}$, J L{\aa}ng$^{4,5}$, P Laukkanen$^{4,5}$, M P J Punkkinen$^{4,5}$ and K Kokko$^{4,5}$} 
\address{$^1$ IFW Dresden, P.O. Box 270116, D-01171 Dresden, Germany}
\address{$^2$ Institute of Physics, Martin Luther University Halle-Wittenberg, D-06099 Halle, Germany}
\address{$^3$ Max Planck Institute of Microstructure Physics, Weinberg 2, D-06120 Halle}
\address{$^4$ Department of Physics and Astronomy, University of Turku, FIN-20014 Turku, Finland}
\address{$^5$ Turku University Centre for Materials and Surfaces (MatSurf), Turku, Finland}
\ead{mart.hoffi@gmail.com}

\begin{abstract}
We investigate the effect of short-range order (SRO) on the electronic structure in alloys
from the theoretical point of view
using density of states (DOS) data. In particular, the interaction 
between the atoms at different lattice sites is affected by chemical disorder, 
which in turn is reflected in the fine structure of the DOS
and, hence, in the outcome of spectroscopic measurements.
We aim \add{at quantifying} the degree of potential SRO with a 
proper parameter.

The theoretical modeling is done with the Korringa-Kohn-Rostoker Green's function 
method. Therein, the extended multi-sublattice non-local coherent 
potential approximation is used to include SRO. As a model system, we use the binary solid 
solution \agpd at three representative concentrations $c=0.25,0.5$ and $0.75$.
The degree of SRO is varied from local ordering to local segregation through an intermediate completely uncorrelated state.
We observe some pronounced features, which change over the whole energy range of the valence bands 
as a function of SRO in the alloy. 
These spectral variations should be traceable in modern photoemission experiments.
\end{abstract}

\noindent\textit{Short-range order, first-principles calculations, AgPd, solid solution, density of states, KKR, HUTSEPOT, photoemission spectroscopy}
\maketitle
\ioptwocol

\begin{figure*}[bhtp]
  \includegraphics[width=\textwidth]{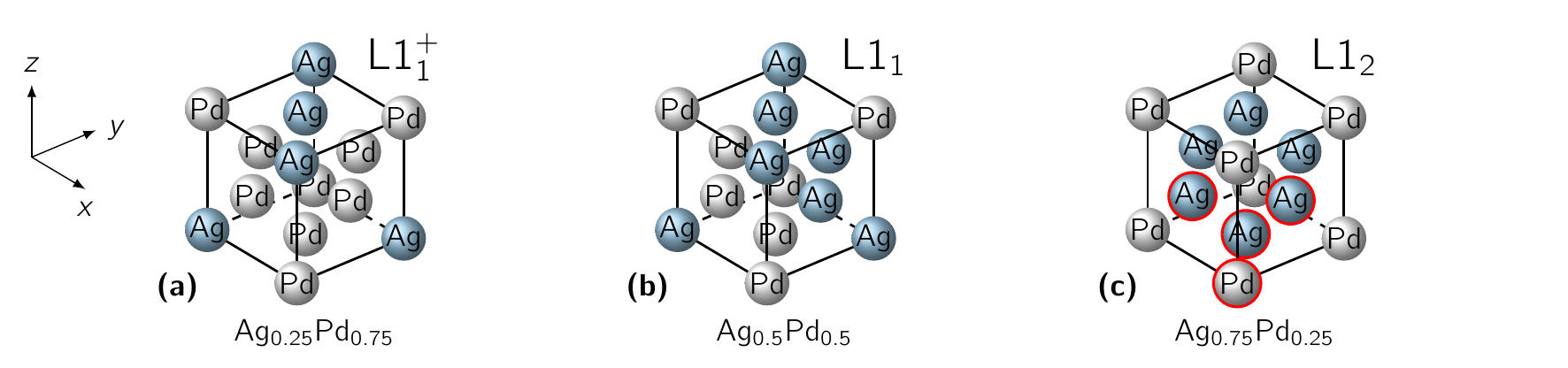}
  \caption{\add{Schematic pictures of the three ordered structures found theoretically by M\"uller and Zunger \cite{Mueller2001}
    for the Ag concentration $c=0.25$, 0.5, and 0.75 ((a), (b) and (c), respectively). Only (c) depicts a unit cell (red circles 
    indicate the basis atoms). L$1_1^+$ is formed of a Pd and a mixed Pd/Ag layer
    in [111] direction, whereas in L$1_1$, a Pd layer  alternates with a full Ag layer in [111] direction.}}
  \label{fig:ordered_structures}
\end{figure*}

\section{Introduction}
Short-range order (SRO), i.e., a partial degree of order within length scales comparable to interatomic distances, 
affects materials properties in many macroscopic ways. Its effects can be 
found in the optical conductivity and reflectivity \cite{Mookerjee2008,Jezierski1993}, magnetism 
\cite{Parra1999}, plasticity \cite{Jiang2010}, and electronic structure
\cite{Jezierski1993,Staunton1994,Wahrenberg2000,Golovchak2011,Novikov2011}.
A systematic study of SRO in \agpd alloys is particularly informative, 
because their experimental phase diagram shows continuous solid solubility within the whole concentration range $c\in[0,1]$, in
a randomly substitutional face centered cubic (fcc) structure
\cite{Hultgren1973}.
Several theoretical predictions of stable long-range order (LRO), i.e., 
perfectly periodic, lower energy phases, have been also made in this system 
\cite{Mueller2001,Ruban2007}. 
These include in particular unit cell types L$1_2$, L$1_1$ and so-called L$1_1^+$ 
(a variation of the L$1_1$ case, with its original Ag layer hosting 
\SI{50}{\percent} Pd atoms\cite{Mueller2001}) at $c=0.75$, $c=0.5$ and $c=0.25$, 
respectively \add{(see \fref{fig:ordered_structures})}.

\add{The present study adds to our previous qualitative investigation of 
SRO effects on elastic and Fermi surface properties of \agpd \cite{Hoffmann2012own} a quantitative
analysis of various degrees of SRO in \agpd. The theoretical approach used in \cite{Hoffmann2012own}
-- the multi-sublattice extension of the dynamical cluster 
approximation \cite{Jarrell2001}
/  non-local coherent potential approximation (MS-NL-CPA) 
\cite{Rowlands2006,Marmodoro2013} -- 
 extends the original single-site coherent potential 
approximation (CPA) allowing} the evaluation of SRO modeled as local environments 
up to a given ``cavity'' size $N_c \times \Nsub$. Here, $\Nsub$ is the number of 
sublattices in a reference unit cell while $N_c$ counts multiple instances
of that unit cell (so-called reciprocal space ``tiles'').
The SRO character of this approach is then included via the variation of 
a possible occupation of the sublattices by alternative atomic species. It is  
typically described through the introduction of a order parameter. One 
example of a SRO parameter, $\alpha$, is offered by the Warren-Cowley definition \cite{Cowley1950pr, Warren1990}, 
which has 
been previously used for proof-of-concept evaluations of SRO effects in 
CuZn alloys \cite{Rowlands2006}, in comparison with actual neutron scattering 
experiments on $\beta$ brass \cite{Walker1963}, and in the first-principles 
study of electrical conductivity \cite{Butler1984,Lowitzer2010}.
Temmermann \etal \cite{Temmermann1988} found from theoretical calculations that
the order-disorder transformation in $\beta$ brass should be visible in
photoemission spectra.

In this work, we further develop such parametrization of the SRO
and target a more quantitative comparison with past 
experiments on \agpd alloys. These alloys are 
on the one hand easy to handle model systems, since they show intermixing
at variable concentrations, but might also stabilize in various geometrically periodic 
yet substitutionally disordered phases. On the other hand, 
practical reasons of interest for such compounds are given by possible 
application for fuel cells, catalysts, hydrogenation, sensors and 
biosensors and dental implantology
\cite{Li2010}. Besides bulk properties, other areas of current interest entail the structure of
Pd-Ag nanoparticles (see Ref. \cite{Kozlov2015} and references therein).

We calculated the total density of states (DOS) for the different
SRO settings and compared the predicted SRO changes with available 
experimental photoelectron spectroscopy (PES) data. Much more features, 
which would allow a specific fingerprint of a
SRO scenario, were visible in the theoretical prediction than in the 
measured PES. Since the available experimental material is quite old and
the present day high-resolution PES methods will allow better differentiation,
we expect that changes in SRO will be traceable within PES experiments.

In the following \sref{sec:comp_details}, 
we describe the adopted theoretical method in its essential details.
The formulation of a suitably general SRO parameter is given in \sref{sec: SRO parameter}.
We use it in \sref{sec:DOS_comparison} to compare between theoretical DOS results at different ordering 
regimes and the
experimental peak positions from PES. 
Our conclusions are summarized in \sref{sec:conclusions}.

\section{Computational details}
\label{sec:comp_details}

\begin{figure*}
  \includegraphics[width=\textwidth]{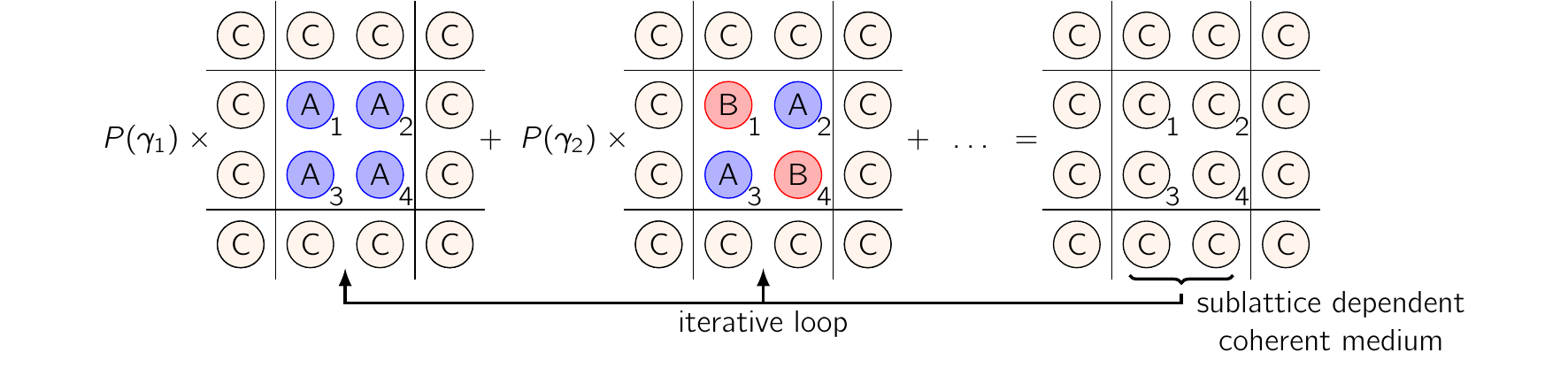}
  \caption{Schematic two-dimensional view of the MS-NL-CPA method
    for $N_\text{sub}=4$ and $N_c=1$.
    Numbers 1 to 4 mark the sublattice positions $s$.}
  \label{fig:MSNLCPA_sketch}
\end{figure*}

The electronic structure calculation scheme of choice
was the Korringa-Kohn-Rostoker Green's function (KKR-GF) method.
Here, the HUTSEPOT code developed by A. Ernst \etal \cite{Ernst2007habil, Luders2001jpcm}
was used.
We adopted the same calculation settings as used in our previous 
work \cite{Hoffmann2012own}.
Thus, the full charge density approximation (FCDA) was applied in 
order to describe the potentials properly
and the local density approximation (LDA) \cite{Perdew1992a}
was used as exchange-correlation functional.
The expansion cut-off for the spherical harmonics in the KKR-GF
was set to $l_\text{max}=3$.
Relaxed lattice parameters as a function of the concentration have been
computed from total energy minimization through a fit to the Birch-Murnaghan
equation of states \cite{Birch1947pr, Poirier2000book}
(\tref{tab:lat_constants}).

\begin{table}
\caption{Calculated equilibrium lattice constants corresponding to the different alloy concentrations.
         The values are taken from CPA results \cite{Hoffmann2012own}.}
\label{tab:lat_constants}
  \lineup
  \begin{indented}
  \item[]
  \begin{center}
  \begin{tabular*}{0.8\columnwidth}{c@{\extracolsep{\fill}}ccc}
    \br
    $c$                             & 0.25  & 0.5   & 0.75  \\
    $a_\text{lat}$ [\si{\angstrom}] & 3.890 & 3.929 & 3.970 \\
    \br
 \end{tabular*}
  \end{center}
 \end{indented}
\end{table}

Within the MS-NL-CPA framework, we describe the SRO considering a multi-site 
cavity, here set up with $N_c=1$ tiles but $\Nsub\geq 1$ sublattices
\cite{Marmodoro2013}. This situation is sketched in \fref{fig:MSNLCPA_sketch}.
Beginning from a starting assumption for the coherent medium,
the calculation is iterated until self-consistency of the coherent medium.
In general, if each disordered cavity site $s$ can host $N_\text{a}(s)$
alternative atomic species, there will be in total
$N_\text{tot}= \prod_{s=1}^{\Nsub} N_a^{N_c}(s)$ 
possible local configurations $\gamma$, each with weight $P(\gamma)$.

This framework allows to recover LRO results when only one, periodically 
repeated configuration occurs with probability one.
On the opposite end, we obtain the fully uncorrelated scenario of a perfectly disordered
lattice (which corresponds to the original single-site CPA picture)
when all $\gamma$ are sampled with a probability distribution
\begin{equation}
  P(\gamma) = \prod_{I=1}^{N_c} \prod_{s=1}^{\Nsub} c_{A(I,s,\gamma)}\mathcomma
  \label{eq:probability_alpha0_MS-NL-CPA}
\end{equation}
only given by the factorized concentrations $c_{A(I,s,\gamma)}$.
They represent the single-site concentration
of an atomic species $A$ appearing on the (MS-)NL-CPA tile 
$I\in \lbrace 1,\ldots,N_c \rbrace$ and the sublattice 
$s\in \lbrace 1,\ldots,\Nsub \rbrace$, when the cavity 
is populated by configuration $\gamma$. Intermediate scenarios 
can be described adopting alternative probability values 
$P(\gamma_i) \in [0,1]$, which are subject to the normalization constraint
\begin{equation}
  \sum_{i=1}^{N_\text{tot}} P(\gamma_i) = 1\mathcomma \label{eq:restriction_Pgamma=1}
\end{equation}
and satisfying the stoichiometry requirement for any atomic type $A$
\begin{equation}
  \frac{1}{N_c \Nsub} \sum_{i=1}^{N_\text{tot}} P(\gamma_i) \times \op{N}_A(\gamma_i) = c_A
  \label{eq:restriction_Pgamma_opp=c_A}
  \mathperiod
\end{equation}
Therein, the factor $\op{N}_A(\gamma_i)$ counts how many atoms of type $A$
appear within configuration $\gamma_i$. 

We note that our results represent an upper limit for the influence 
of SRO effects on the physical effective medium, 
on top of those due to concentration alone. This originates 
from the $N_c$ coarse-graining subdivisions of the original Brillouin zone
in reciprocal space \cite{Rowlands2008}, which 
are chosen consistently with the point group symmetries of the lattice
but remain only defined up to a systematic offset 
(or ``tiling phase factor'' \cite{Rowlands2008}) in the relative
origin for the cluster momenta $\vec{K}_n$.
At the moment, there is no systematic KKR-GF implementation of a corrective,
additional sampling step for the tiling phase factor. 
Therefore, single tiling phase results - such as those discussed in 
the following - might slightly broaden, when including a 
proper phase average.

\subsection{A general short-range order parameter}
\label{sec: SRO parameter}

We intend to \add{improve our previous work \cite{Hoffmann2012own}
with} a quantitative description of SRO, and facilitate a comparison
with experiments.
To this end, we begin by recalling the Warren-Cowley SRO parameter 
definition \cite{Cowley1950pr, Warren1990}.
It is for a generic $A_{c},B_{1-c}$ binary alloy
computed from the number $n_l$ of $A$ atoms found 
in the $l$-th shell around a $B$ atom
\begin{eqnarray}
  \alpha_l^{BA} & = 1-\frac{n_l}{c_A C_l}  \label{eq:SRO_parameter_Warren_Cowley}\\
                & = 1-\frac{p^{BA}_l}{c_A}  \label{eq:SRO_parameter_Warren_Cowley_probability}
  \mathcomma
\end{eqnarray}
Therein, $C_l$ is the coordination number of the $l$-th shell
around $B$ and the second expression 
is obtained by inserting the ratio  
$p^{BA}_l={n_l}/{C_l}$ of $A$ 
atoms within shell $l$ around a $B$ atom.
Complex unit cell cases can be handled through an additional 
$\Nsub$-normalized summation across sublattices. 

When considering a (MS-)NL-CPA cavity,
the SRO parameter 
\eref{eq:SRO_parameter_Warren_Cowley}
is deployed for each configuration $\gamma_i$, 
leading to the global result as a $P(\gamma_i)$-weighted average.
This general case can present some difficulties,
since the shell radius in \eref{eq:SRO_parameter_Warren_Cowley}
may exceed the cavity size, so that the 
occupation of the considered lattice sites lies
beyond the explicit listing of a configuration $\gamma_i$
(see \fref{fig:MSNLCPA_sketch}, colored spheres are ``inside'' and C spheres are ``outside''
of the cavity).
We propose therefore a general SRO parameter defined by the procedure below.
Therein, it is convenient to introduce a generic
occupation function $\occ{s}{A}$ for each atomic species $A$, 
which returns a value ``1'', if the crystalline position $s$
under examination hosts an $A$ atom, or ``0'', if not.
At every instance, an example is given with respect to 
the configuration $\gamma_2$ in \fref{fig:MSNLCPA_sketch} (middle panel)
with $\Nsub=4$ and $N_c=1$ (see also \sref{sec:nsub4}).

\begin{enumerate}
  \item As long as a sublattice remains fully contained within the explicit configuration, 
        the constrained probability for $A$ to appear 
        on sublattice $s$ of the (MS-)NL-CPA cavity cell $I$ is
        given by the occupation function
        $\occ{I,s}{A}$.
        When instead a shell's site lies beyond the cavity, its constrained
        probability becomes
        \begin{eqnarray}
          p^{A}_\text{C}(I,s) & = \sum_{i=1}^{N_\text{tot}}  P(\gamma_i) \occ{I,s}{A}
          \label{eq:coherent_probability}
          \mathperiod
        \end{eqnarray}
        We note that in case of a disordered site, in the sense of 
        the single-site CPA, $N_\text{tot}=1$ and the occupation function 
        in \eref{eq:coherent_probability} is substituted by a concentration.
    
        \textbf{Example:} Sites 1, 2, 3, and 4 are determined by $\gamma_2$ (inside the cavity),
        while all sites marked with C are not included
        in any configuration (outside the cavity).

  \item The conditional probability in \eref{eq:SRO_parameter_Warren_Cowley_probability}
        for each configuration $\gamma_i$ and across the shell $l$ is computed by
        \begin{eqnarray}
          p^{BA}_l(\gamma_i,I,s) = \frac{1}{C_l} 
                      \sum_{n=1}^{C_l} 
                        \cases{ {p}_{A}(I,s) & \text{$n$ inside,} \\
                                \occ{I,s}{A}       & \text{$n$ outside,} \\ }
          \label{eq:prob_in_out}
        \end{eqnarray}
        where the $n=1,\ldots,C_l$ sites are
        either inside or outside the cavity.
        
        \textbf{Example:} Site 1 has four nearest neighbor sites. 
        The probability of sites 2 and 3
        is taken into account within $\gamma_2$ by 
        $\occ{1,2}{A}$ and $\occ{1,3}{A}$, respectively,
        whereas the other two sites are outside and their
        probability is $p^{A}_\text{C}(1,2)$ and $p^{A}_\text{C}(1,3)$.
  \item Using \eref{eq:prob_in_out} in 
        \eref{eq:SRO_parameter_Warren_Cowley_probability} yields the 
        SRO parameter $\bar{\alpha}_l(\gamma_i)$  as a function of 
        the configuration $\gamma_i$.
        The arithmetic average is taken over all sublattices 
        and tiles for the $l$-th shell 
        (using $\alpha^{AB}_l(\gamma_i,I,s)$ or
        $\alpha^{BA}_l(\gamma_i,I,s)$).
        
        \textbf{Example:} Average over the sites 1, 2, 3 and 4.
  \item In a final step, the SRO parameter per shell is derived via
        \begin{eqnarray}
          \alpha_l = \sum_{i=1}^{N_\mathrm{tot}}  P(\gamma_i)\times\bar{\alpha}_l(\gamma_i)
          \mathperiod
          \label{eq:exact SRO parameter Warren Cowley}
        \end{eqnarray} 

        \textbf{Example:} Take into account all other configurations and the corresponding
        $P(\gamma_i)$ as well.
  \item A possible average over the shells up to $N_\text{sh}$ may include weighting with the coordination number $C_l$ 
        \cite{Mirebeau1984prl}
	\begin{eqnarray}
	  \langle\alpha\rangle_{N_\text{sh}} = \Big[{\sum_{l=1}^{N_\text{sh}}  C_l\times\alpha_l}\Big]\Big/\sum_{l=1}^{N_\text{sh}}  C_l
	  \mathperiod
	\end{eqnarray}
	However, $N_\text{sh}$ is not yet defined. Its value may depend on the lattice structure as discussed 
	below in \sref{sec:varying_SRO_parameter}. Thus, we restrict this study to the 
	nearest neighbor SRO parameter. 
\end{enumerate}
\begin{figure}
  \includegraphics[width=\columnwidth]{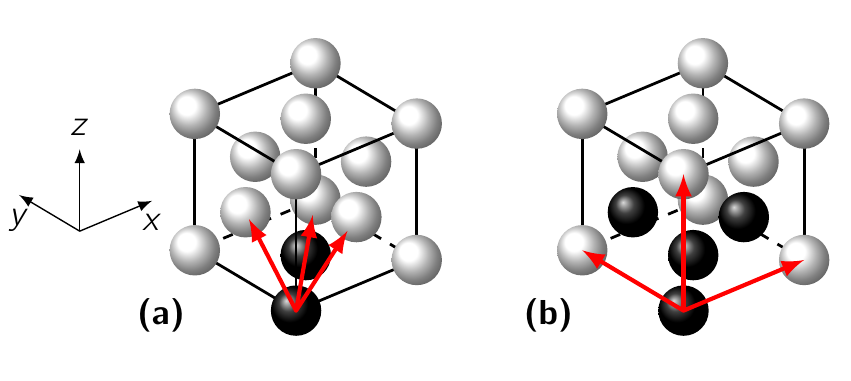}
  \caption{\add{The used cells for the representation of the fcc lattice. (a) with two basis sites.
     (b) with 4 basis sites. Red arrows indicate the lattice vectors,
     and black spheres represent the basis sites.}}
  \label{fig:lattice_structures}
\end{figure}

\subsection{Parameter space of the probabilities: an example}
\label{sec:SRO_example_Nsub2}
Although the definition of the SRO parameter consists 
only of different kinds of averages, the choice of the probabilities
is still an open question. The connection of the SRO parameter 
with the probabilities can be only visualized for
a simple test case, which contains only $\Nsub=2$ sublattices
with $N_\text{tot}=2^2=4$ possible configurations. Otherwise, 
the number of configurations becomes too large. 
Therefore, we begin at first to consider 
this $\Nsub=2$ example for a 
lattice, which is described by the vectors 
$\vec{R}_1=(0, 1/2, 1/2)$, $\vec{R}_2=(1/2, 0, 1/2)$ and
$\vec{R}_3=(1, 1, 0)$ and 
the basis vectors $\vec{a}_1=(0,0,0)$ and $\vec{a}_2=(1/2,1/2,0)$
\add{(see \fref{fig:lattice_structures}a)}.
This lattice structure resembles an fcc lattice.

\begin{figure}
  \includegraphics[width=\columnwidth]{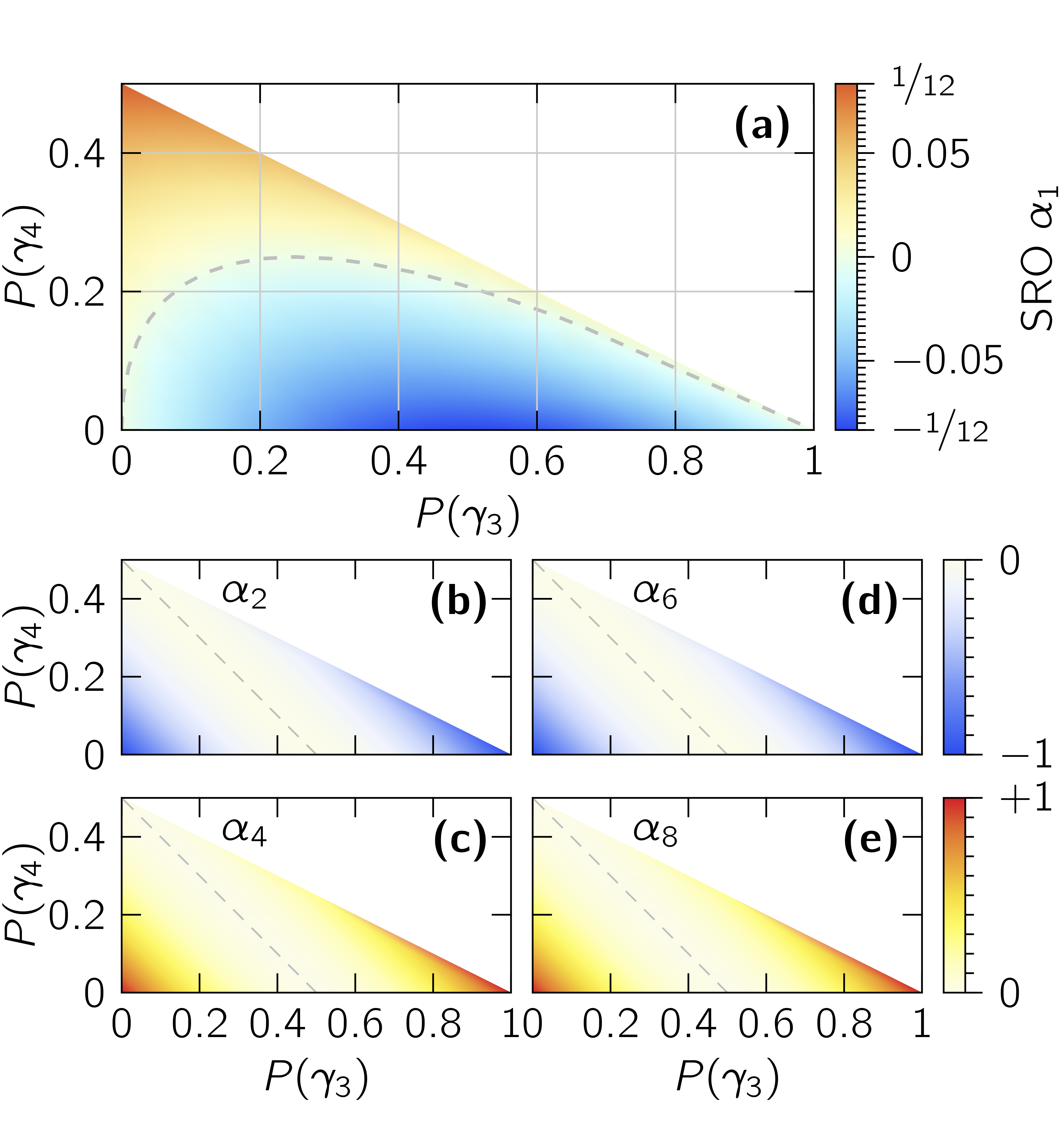}
  \caption{(a) The nearest neighbor SRO parameter $\alpha_1$,
      color coded as a function of 
      the two probabilities $P(\gamma_3)=P(\text{Pd\,Ag})$ and
      $P(\gamma_4)=P(\text{Pd\,Pd})$. The triangle shape 
      follows from the restrictions in 
      \eref{eq:Nsub probabilities equation system gamma_3}
      and \eref{eq:Nsub probabilities equation system gamma_4}.
      The dashed line indicates $\alpha_1=0$. 
      (b)-(e) The SRO parameter for the following shells
      vary between $-1$ and $1$ and repeat themselves (see text). 
      The SRO parameter not shown 
      ($\alpha_3$, $\alpha_5$, $\alpha_7$) are completely zero.
      }
  \label{fig:SRO_parameter_alpha_Nsub2}
\end{figure}

The system of equations formed from \eref{eq:restriction_Pgamma=1}
and \eref{eq:restriction_Pgamma_opp=c_A}
has only one solution and yields for the four probabilities
\begin{eqnarray}
  P(\gamma_1) = P(\gamma_4) \mathcomma\label{eq:Nsub probabilities equation system gamma_1}\\
  P(\gamma_2) = 1 -  P(\gamma_3) - 2  P(\gamma_4)\mathcomma\label{eq:Nsub probabilities equation system gamma_2}\\
  0\leq P(\gamma_3) \leq 1 \mathcomma\label{eq:Nsub probabilities equation system gamma_3}\\
  0\leq P(\gamma_4)\leq \frac{1-P(\gamma_3)}{2}\mathperiod\label{eq:Nsub probabilities equation system gamma_4}
\end{eqnarray}
The two latter probabilities are free parameters. These allow 
a graphical analysis of the SRO parameter in a contour plot
(see \fref{fig:SRO_parameter_alpha_Nsub2}). The local variations inside the 
cavity determines only the nearest neighbor 
SRO parameter $\alpha_1$ (see \fref{fig:SRO_parameter_alpha_Nsub2}a). 
It varies between $-\nicefrac{1}{12}$ and $+\nicefrac{1}{12}$.
The highest degree of order is found for $P(\gamma_3)=0.5$ and 
$P(\add{\gamma_4})=0$, which means having the configurations (Pd Ag) and
(Ag Pd) equally distributed. On the other hand, the highest degree
of segregation in $\alpha_1$ is realized having $P(\gamma_1)=P(\add{\gamma_4})=0.5$.

The higher shells reflect the periodicity of the 
underlying lattice and the coherent medium
(see \fref{fig:SRO_parameter_alpha_Nsub2}b to 
\ref{fig:SRO_parameter_alpha_Nsub2}e). 
Due to the choice of the lattice
and basis vectors, the first period includes the shells until $l=5$.
However, we restrict in this study the average of the SRO parameter
to the non-periodic contribution, since this
represents mainly the character of the SRO.

\subsection{A reasonable choice of sublattices}
\label{sec:nsub4}

Although the example demonstrates well the 
concept of the SRO parameter in the MS-NL-CPA,
its configuration space is a little bit too restricted. 
Therefore, we considered a $\Nsub=4$
sublattice supercell \add{sketched in \fref{fig:lattice_structures}b}
(lattice vectors of a simple cubic cell with the basis of
$\vec{a}_1=(0,0,0)$, $\vec{a}_2=(1/2,1/2,0)$,
$\vec{a}_3=(1/2,0,1/2)$
and $\vec{a}_4=(0,1/2,1/2)$). 
In this case, the corresponding probabilities of the
16 potential configurations can not be parametrized by two 
free values. 

\begin{table}
  \caption{All 16 possibilities for the occupation of $\Nsub=4$ sublattices in {fcc} {\agpd}.
           The last column defines new probabilities $\tilde{P}(\mathcal{N}_\text{Ag}(\gamma_i))$.}
  \label{tab:Nsub=4_configurations}
  \lineup
  \begin{indented}
  \item[]
  \begin{center}
  \begin{tabular*}{\columnwidth}{@{\extracolsep{\fill}}ccccc|ccc|c }
    \br
     & \multicolumn{4}{c|}{configurations $\gamma_i$} & & & \\
    $i$ & $\vec{a}_1$ & $\vec{a}_2$ & $\vec{a}_3$ & $\vec{a}_4$ & $c_\text{Ag}$ & $c_\text{Pd}$ & $\mathcal{N}_\text{Ag}(\gamma_i)$ & \\ 
    \mr
 1 & Ag & Ag & Ag & Ag & 1 & 0 & 4 & $\tilde{P}(4)$ \\
    \mr
 2 & Ag & Ag & Ag & Pd & $\tquarter$ & $\quarter$ & 3 &$\tilde{P}(3)$\\
 3 & Ag & Ag & Pd & Ag & $\tquarter$ & $\quarter$ & 3 &\\
 4 & Ag & Pd & Ag & Ag & $\tquarter$ & $\quarter$ & 3 &\\
 5 & Pd & Ag & Ag & Ag & $\tquarter$ & $\quarter$ & 3 &\\
    \mr
 6 & Ag & Ag & Pd & Pd & $\half$ & $\half$ &  2 &$\tilde{P}(2)$\\
 7 & Ag & Pd & Ag & Pd & $\half$ & $\half$ &  2 &\\
 8 & Ag & Pd & Pd & Ag & $\half$ & $\half$ &  2 &\\
 9 & Pd & Ag & Ag & Pd & $\half$ & $\half$ &  2 &\\
 10 & Pd & Ag & Pd & Ag & $\half$ & $\half$ & 2 & \\
 11 & Pd & Pd & Ag & Ag & $\half$ & $\half$ & 2 & \\
    \mr
 12 & Ag & Pd & Pd & Pd & $\quarter$ & $\tquarter$ & 1 & $\tilde{P}(1)$\\
 13 & Pd & Ag & Pd & Pd & $\quarter$ & $\tquarter$ & 1 & \\
 14 & Pd & Pd & Ag & Pd & $\quarter$ & $\tquarter$ & 1 & \\
 15 & Pd & Pd & Pd & Ag & $\quarter$ & $\tquarter$ & 1 & \\
    \mr
 16 & Pd & Pd & Pd & Pd & 0 & 1 & 0 & $\tilde{P}(0)$\\
    \br
  \end{tabular*}
  \end{center}
  \end{indented}
\end{table}

However, the restrictions in \eref{eq:restriction_Pgamma=1} and 
\eref{eq:restriction_Pgamma_opp=c_A}
depend only on the number of $A$ or $B$ types in each configuration
(internal concentration), whereby several configurations 
have an equal number of atomic types, which occupy only different sublattices
(see \tref{tab:Nsub=4_configurations}). A new probability $\tilde{P}(\mathcal{N}_\text{Ag}(\gamma_i))$ 
is assigned to every group depending on the number of Ag atoms $\mathcal{N}_\text{Ag}(\gamma_i)$.
With these 5 probabilities, the system of equations \eref{eq:restriction_Pgamma=1}
and \eref{eq:restriction_Pgamma_opp=c_A}
can be solved again, where two probabilities are determined by the others
(see \ref{sec:appedix_probability_relations}).
In fact, each $\tilde{P}(\mathcal{N}_\text{Ag}(\gamma_i))$ describes a subset of 
configurations, e.g., $\tilde{P}(3)$ condenses four configurations ($\gamma_2$ to $\gamma_5$),
each having one Pd occupying another sublattice while the
three sublattices left are occupied with Ag. 
Then, the probabilities $P(\gamma_2)$, $P(\gamma_3)$, $P(\gamma_4)$, and $P(\gamma_5)$ 
are free to choose but have to sum up to $\tilde{P}(3)$,
otherwise violating the total concentration.

\subsection{Comparison with experimental PES}
We compare below our calculations of the 
DOS for
different SRO regimes
with experimental valence band PES of Ag-Pd alloys by McLachlan \etal \cite{McLachlan1975}.
The mean positions of the experimentally observed spectral peaks are considered as the 
electron binding energies and are given in \tref{tab:exp_binding_energies_HeII}.
We preferred in particular
the experimental He II (\SI{40.81}{\electronvolt}) spectra.
Although the He II 
technique is in general rather surface sensitive, 
we expect that its application to metals with a highly efficient electronic screening
can lead to useful insights on the bulk properties from analysis of the spectra.
This is further confirmed by comparison of the specific
He II results used in this study against
the calculated XPS spectra of Winter \etal \cite{Winter1984}, and the typical probing depth 
of about $\SI{50}{\angstrom}$ reported in experiments by Caroli \etal \cite{Caroli1973prb}, thus
including substantial bulk contributions. 

\begin{table}
  \caption{Binding energies (in $\si{\electronvolt}$, relative to the Fermi energy) of the main spectral 
  peaks estimated from the experimental (He II) spectra by McLachlan \etal \cite{McLachlan1975} 
  and interpolated to the concentrations used in the present calculations. 
  The upper two rows are 
  for the Pd $4d$ section of the spectrum and the lower two rows belong to the Ag $4d$ section.}
  \label{tab:exp_binding_energies_HeII}
  \lineup
  \begin{indented}
  \item[]
  \begin{center}
  \begin{tabular*}{0.8\columnwidth}{c@{\extracolsep{\fill}}cc} \\
    \br
    \multicolumn{1}{c}{\agpdc{0.25}{0.75}} &\multicolumn{1}{c}{ \agpdc{0.50}{0.50}} & \multicolumn{1}{c}{\agpdc{0.75}{0.25}} \\ \mr
    $0.5$ & $1.0$ & $1.4$ \\
    $2.4$ & $2.3$ & $-$   \\
    $4.9$ & $4.6$ & $4.4$ \\
    $5.5$ & $5.7$ & $6.0$ \\
    \br
  \end{tabular*}
  \end{center}
  \end{indented}
\end{table}

\section{Results}
\label{sec:DOS_comparison}
\subsection{Broadening of the theoretical spectrum}
\label{sec:broadening}

When comparing theoretical DOS data with experimental results, the experimental resolution 
broadens the measured spectra and may hide some spectral 
features. 
The experimental resolution in the study
of McLachlan \etal \cite{McLachlan1975} is given by $\pm\SI{0.3}{\electronvolt}$.
However, in the modern high-resolution photoelectron measurement equipment, 
the energy resolution can go down to the range of few \si{\milli\electronvolt}
at low temperatures around \SI{10}{\kelvin} \cite{Hufner1999, Stadnik2001prb}.
The influence of the experimental resolution on the calculated DOS can
be simulated by the convolution of the DOS with a Gaussian. 
The resolution is understood as the full width at half maximum (FWHM)
and is translated to the standard deviation $\sigma$ of the 
Gaussian distribution by $\text{FWHM}=2\sqrt{2\ln{2}}\sigma$. 

\begin{figure}[t]
  \includegraphics[width=\columnwidth]{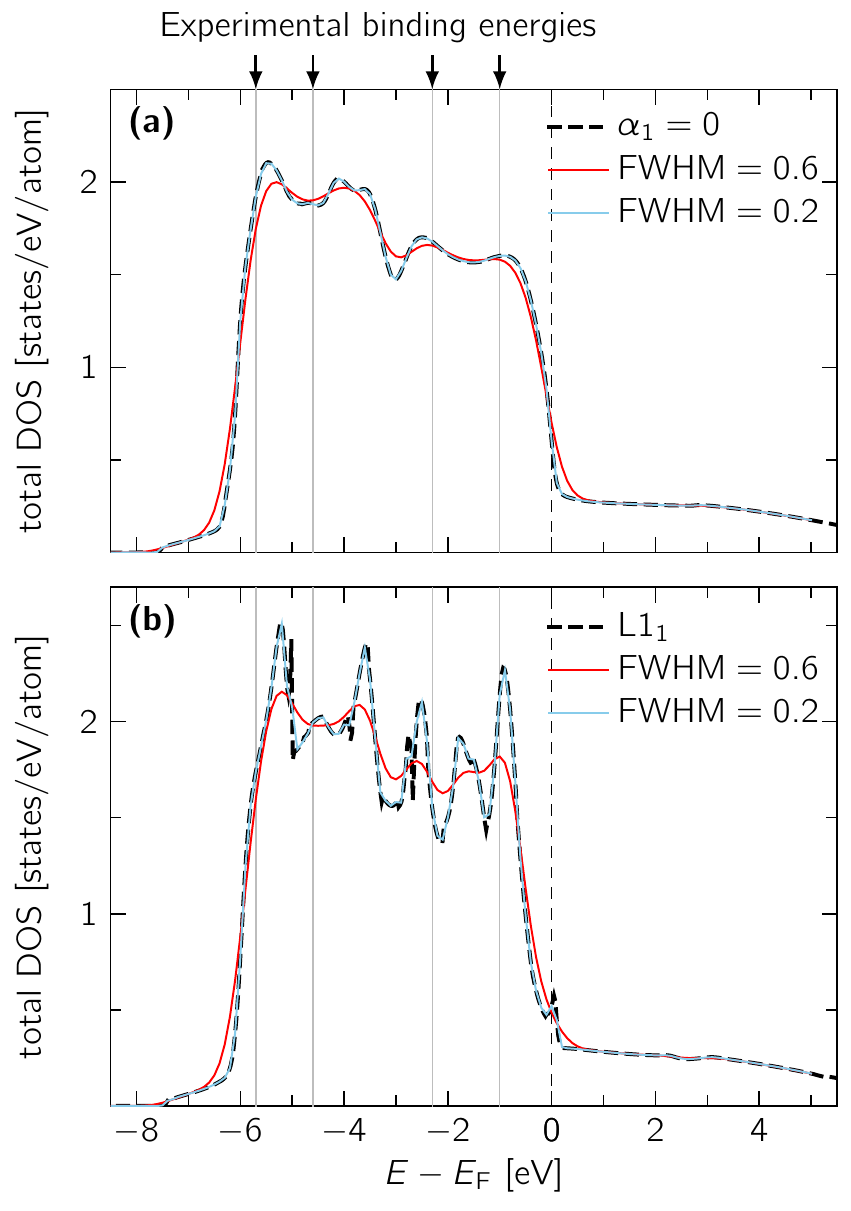}
  \caption{Calculated density of states of \agpdc{0.5}{0.5} convoluted with 
    a Gaussian of different FWHM in order to simulate the experimental
    resolution. Two cases of SRO are depicted (a) $\alpha_1=0$ and (b) the ordered L$1_1$ structure. 
    The binding energies of the main spectral peaks of the experimental PES 
    (\tref{tab:exp_binding_energies_HeII}) 
    are highlighted by arrows and vertical gray lines. 
    The spectra for the high experimental resolution ($\text{FWHM}=0.2$, blue line) and the purely theoretical spectra
    (black dashed line) lie almost on top of each other.} 
  \label{fig:resolution}
\end{figure}

For two extreme SRO regimes, $\alpha_1=0$ (totally uncorrelated) and
the L$1_1$ structure (order), the calculated and broadened DOS are
depicted in \fref{fig:resolution}. The first choice of $\fwhm=0.6$
(red solid lines) corresponds to the resolution of the older experiment
\cite{McLachlan1975}, whereas the $\fwhm=0.2$ (light blue lines)
matches with modern resolutions at room temperature.
Already the latter resolution is sufficient to represent all significant peaks
in the calculated DOS (black dashed line), even for the spiky DOS of the 
ordered structure (see \fref{fig:resolution}b). 
It shows that this resolution would be in principle enough
to differentiate between different SRO regimes with the
combination of first-principles calculations and PES measurements.
This is difficult with the older resolution, since
the number of peaks and their variation is hardly distinguishable for the
two examples $\alpha_1=0$ and L$1_1$ (compare red lines in \fref{fig:resolution}).
A comparison with the experimental peak position (see 
\tref{tab:exp_binding_energies_HeII}) does not reveal a 
clear conclusion about the particular state of order.

\subsection{Varying the short-range order}
\label{sec:varying_SRO_parameter}

\begin{figure}
  \includegraphics[width=\columnwidth]{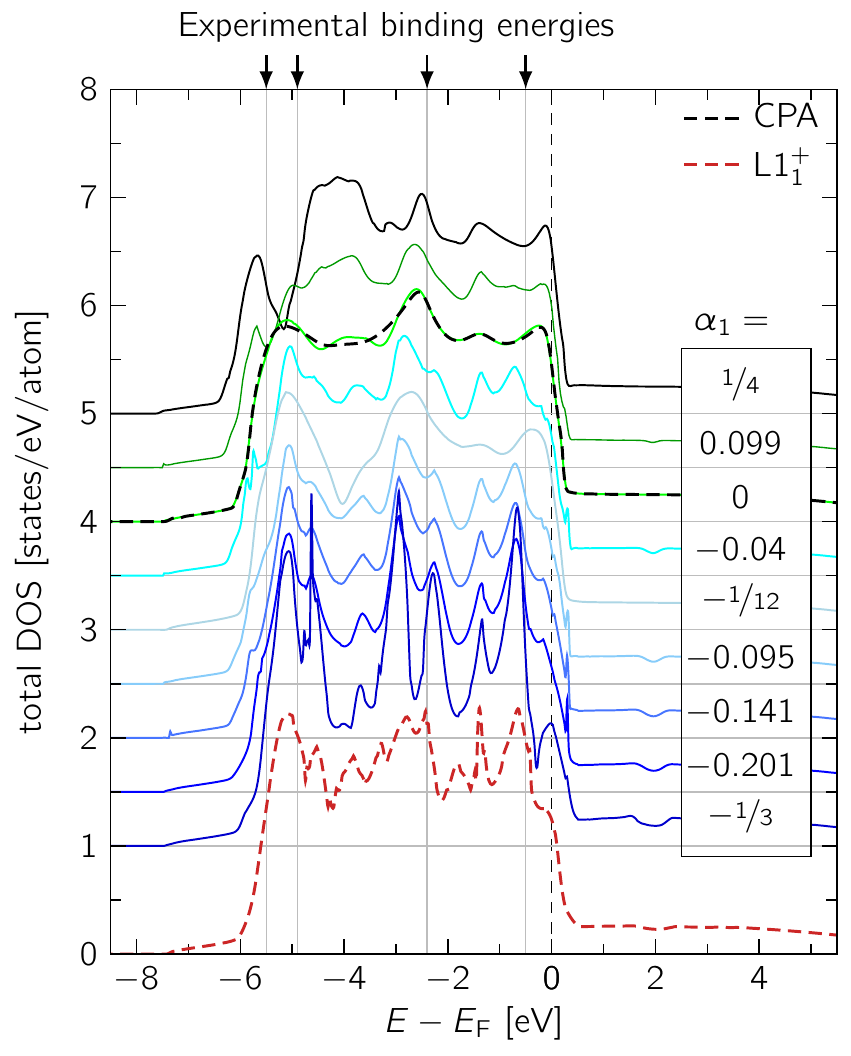}
  \caption{Calculated density of states of \agpdc{0.25}{0.75} for different degrees of nearest 
    neighbor SRO $\alpha_1$,
    beginning with the ordered L$1_1^+$ structure. 
    The corresponding configurations used for the SRO parameter are given in \tref{tab:Nsub=4_configurations_c=0.25}.
    An offset is added to the curves (horizontal gray line represents zero).
    The binding energies of the main spectral peaks of the experimental PES 
    (\tref{tab:exp_binding_energies_HeII}) 
    are highlighted by arrows and vertical gray lines.} 
  \label{fig:c25}
\end{figure}

\begin{figure}
  \includegraphics[width=\columnwidth]{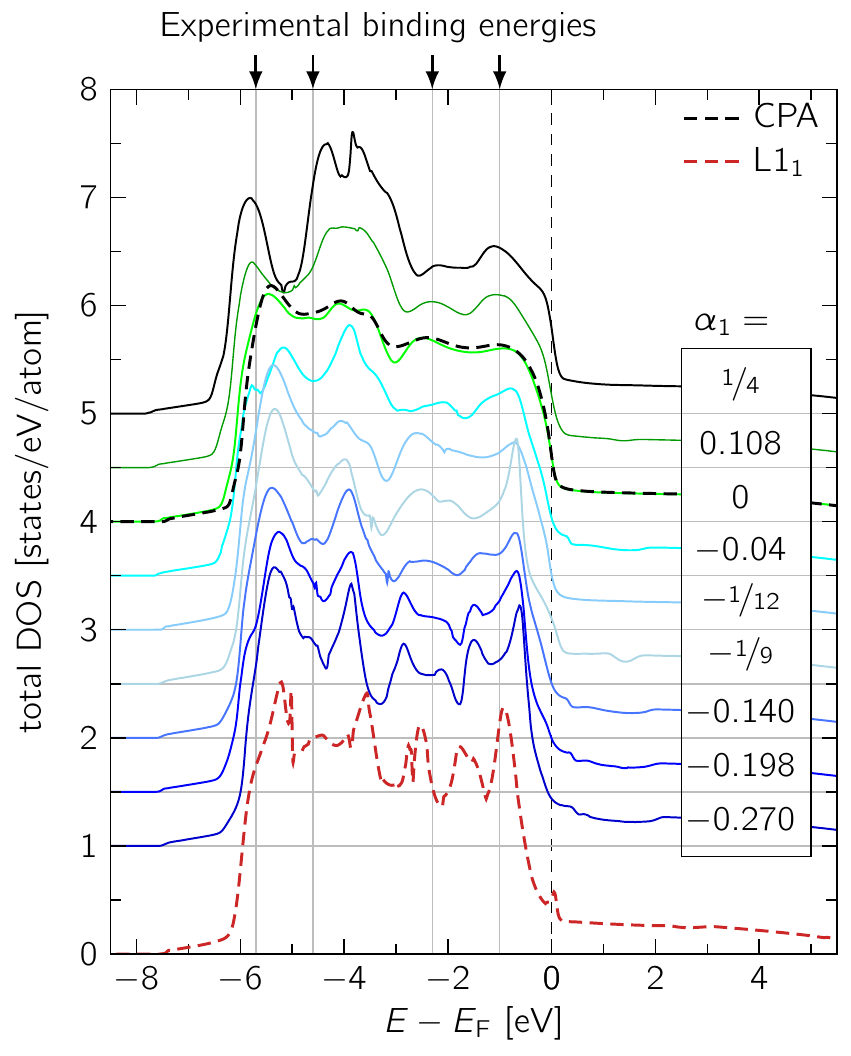}
  \caption{Calculated density of states of \agpdc{0.5}{0.5} for different degrees of nearest 
    neighbor SRO $\alpha_1$,
    beginning with the ordered L$1_1$ structure.
    The corresponding configurations used for the SRO parameter are given in \tref{tab:Nsub=4_configurations_c=0.5}.
    An offset is added to the curves (horizontal gray line represents zero).
    The binding energies of the main spectral peaks of the experimental PES 
    (\tref{tab:exp_binding_energies_HeII}) 
    are highlighted by arrows and vertical gray lines.} 
  \label{fig:c50}
\end{figure}

\begin{figure}
  \includegraphics[width=\columnwidth]{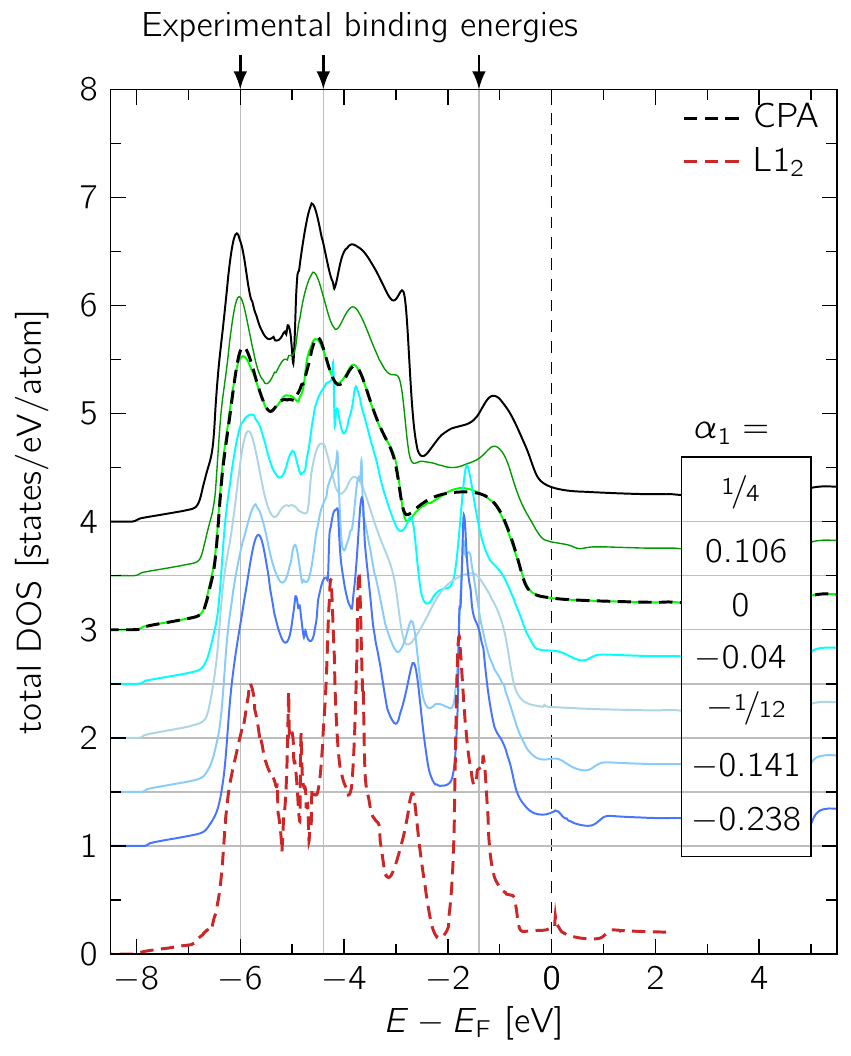}
  \caption{Calculated density of states of \agpdc{0.75}{0.25} for different degrees of nearest 
    neighbor SRO $\alpha_1$,
    beginning with the ordered L$1_2$ structure. 
    The corresponding configurations used for the SRO parameter are given in \tref{tab:Nsub=4_configurations_c=0.75}.
    An offset is added to the curves (horizontal gray line represents zero).
    The binding energies of the main spectral peaks of the experimental PES 
    (\tref{tab:exp_binding_energies_HeII}) 
    are highlighted by arrows and vertical gray lines.} 
  \label{fig:c75}
\end{figure}

As a second step, we varied the degree of SRO at the 
\agpd alloy concentrations $c=0.25$, $0.50$, $0.75$.
We started with only 5 representative configurations of \tref{tab:Nsub=4_configurations},
in particular 1, 2, 6, 12, 16, and 
varied the three probabilities, which are the free parameters (see \ref{sec:appedix_probability_relations}), 
in steps of 0.05. The obtained SRO parameter showed
again a periodicity as already discussed in \sref{sec:SRO_example_Nsub2}. We chose several 
SRO parameter values and calculated the valence DOS.
The results in dependence of the nearest neighbor SRO parameter $\alpha_1$
and for the ordered structures
are depicted in \fref{fig:c25},
\ref{fig:c50} and \ref{fig:c75}, respectively. 
The three figures show significant changes in the DOS with varying SRO. 
Some spectral peaks vanish,
move or grow. Such strong variations should be easily visible in 
nowadays PES measurements. 

When going from the ordered regime ($\alpha_1<0$) via the 
totally uncorrelated case ($\alpha_1=0$) towards the segregation behavior ($\alpha_1>0$),
the spiky structure of the 
DOS looses its contrast and becomes smoother. Simultaneously, 
the band width is enhanced with increasing $\alpha_1$. 
Additionally, the experimental binding energies 
(see \tref{tab:exp_binding_energies_HeII}) are indicated within the 
\fref{fig:c25}, \ref{fig:c50} and \ref{fig:c75} with arrows and thin gray lines. 
Although it became obvious in \sref{sec:broadening} that 
a direct comparison between the experimental and theoretical 
results is hardly possible, the binding energies can 
at least be related with some pronounced peaks in the DOS
and may offer a crude estimation of possible 
SRO scenarios. 

For \agpdc{0.25}{0.75}, the best agreement with
the binding energies would be achieved with the assumption of 
a slight tendency of SRO around $\alpha_1=0$, since the double 
peak structure of the binding energies in the lower energy spectrum
may hint to additional features coming from SRO
(see \fref{fig:c25}).
Nevertheless, the variation 
in the amount of SRO in \agpdc{0.25}{0.75} 
visualizes the gradually collapse or development of 
several spectral peaks when going from negative to positive 
$\alpha_1$. The minimal value for $\alpha_1$ is $-\nicefrac{1}{3}$
and represents again the L$1_2$ structure 
(but now, a Ag atoms at the corner and a Pd atom at the face of the cube). In contrast,
L$1_1^+$ was found to be energetically more favorable
but has a lower degree of ordering in terms of $\alpha_1$
($\alpha_1=-\nicefrac{1}{9}$, $\alpha_2=-\nicefrac{1}{3}$,
$\alpha_3=-\nicefrac{1}{9}$, $\ldots$).
The averaged SRO parameters are
$\langle\alpha\rangle_{2}=-\nicefrac{5}{27}\approx-0.185$ or
$\langle\alpha\rangle_{3}=-\nicefrac{1}{7}\approx-0.143$. 
The different
amount of SRO in L$1_2$ or L$1_1^+$ is directly 
visible in the DOS of both structures \add{(see the dark blue line or the red dashed line in \fref{fig:c25})}. While the DOS 
of L$1_2$ \add{($\alpha_1=-\nicefrac{1}{3}$)} yielded sharper spectral peaks, the DOS 
of L$1_1^+$ \add{matches} better between $\alpha_1=-0.141$
and $\alpha_1=-0.201$.

The analysis of the DOS for \agpdc{0.5}{0.5} is quite similar 
as for \agpdc{0.25}{0.75}. Several spectral peaks become wider and shift
their positions (see \fref{fig:c50}). Also for this concentration,
the proposed ordered structure L$1_1$ has not the lowest possible
SRO parameter (minimum is $\alpha_1=-1$, but for L$1_1$ is 
$\alpha_1=0$, $\alpha_2=-1$,
$\alpha_3=0$, $\alpha_4=1$, $\ldots$).
The DOS does not seem to fit well in respect of the other DOS
of the remaining SRO scenarios. The 
symmetric cubic cell with $\Nsub=4$ might not be 
the best choice of comparing with the layered 
structure of L$1_1$. In terms of the 
experimental binding energies,
the SRO regime of $\alpha_1=0$ agrees
best with the theoretically calculated 
number of spectral peaks and their positions. 

When further raising the concentration of Ag to Ag$_{0.75}$Pd$_{0.25}$, 
the SRO related widening of the spectral peaks observed for the ordered structure
L$1_2$ can be traced
(see energy range between \SIrange{-2}{0}{\electronvolt} in \fref{fig:c75}).
L$1_2$ \add{($\alpha_1=-\nicefrac{1}{3}$)} has already the lowest possible SRO parameter and 
is described well by the small cubic cell. Thus, all spectral peaks
obtained for L$1_2$ just loose their height 
and become broader, if the SRO is varied towards $\alpha_1=0$.
The comparison with the experimental binding energies at $c=0.75$
indicates again a mostly disordered sample representing
the crucial peaks in the theoretical spectrum well
(see arrows in \fref{fig:c75}).

\add{The good description of the $c=0.75$ case within 
the $\Nsub=4$ supercell (see \fref{fig:lattice_structures}) is also verified by calculated
total energies. Thereby, the L$1_2$ structure had the lowest total energy and the total energy 
increased just linearly (not shown) when varying the degree of SRO.
However for the other two concentrations, there was no clear tendency visible.
Only the respective ordered structures -- L$1_1$ and L$1_1^+$ -- had the lowest total energies.}

Finally, the calculated DOS at $c=0.25$, $0.5$ and $0.75$ were also compared
with the PES measurements of Norris and Nilsson \cite{Norris1968}, H\"ufner \etal \cite{Hufner1973,Hufner1975ssc}, 
Chae \etal \cite{Chae1996} and Traditi \etal \cite{Tarditi2012}. 
In general, the experimental spectra agree best with the DOS of the random ($\alpha_1=0$)
or the ordering ($\alpha_1<0$) cases, while the clustering features ($\alpha_1>0$) are less probable. This is in 
agreement with the complete solubility of Ag and Pd at ambient temperatures and with the
ordering tendency at low temperatures \cite{Mueller2001}.

\section{Conclusions}
\label{sec:conclusions}

The SRO induced changes in the DOS are 
significantly larger than the typical energy resolution in the valence band PES 
measurements \cite{McLachlan1975}. We have demonstrated that the SRO phenomena in alloys
can be in principle discernible in valence band photoelectron spectra. 
With proper SRO calculations, e.g., within the MS-NL-CPA, 
the experimental PES data can be used to determine the 
type of the prevailing SRO. Thus,
the PES technique can be considered as one potential 
experimental method to investigate SRO structures of alloys. 

Comparing our MS-NL-CPA valence DOS of Pd-Ag alloys with existing PES measurements
suggests that the SRO in the measured Pd-Ag samples has been in most cases 
that of uncorrelated disorder with some traces of ordering.
Nevertheless, PES measurements with resolution available in modern technique would 
be beneficial to get more definite information of SRO in Pd-Ag alloys.

\section*{Acknowledgments}
This work was partially funded by the Deutsche Forschungsgemeinschaft (DFG) within SFB 
762, “Functionality of Oxide Interfaces.” We gratefully acknowledge financial
support by the Deutscher Akademischer Austauschdienst (DAAD) and the Academy
of Finland (Grant No. 57071667).

\appendix

\section{Relation between probabilities}
\label{sec:appedix_probability_relations}

The redefined probabilities $\tilde{P}(\mathcal{N}_\text{Ag}(\gamma_i))$ form 
a similar system of equations as \eref{eq:restriction_Pgamma=1} and
\eref{eq:restriction_Pgamma_opp=c_A}
\begin{eqnarray}
  \tilde{P}(4) + \nicefrac{3}{4} \tilde{P}(3) +  \nicefrac{1}{2} \tilde{P}(2) + \quarter \tilde{P}(1)  & = c
  \mathcomma
  \label{eq:probability_equation_system1}
\end{eqnarray}
\begin{eqnarray}
  \quarter \tilde{P}(3) +  \nicefrac{1}{2} \tilde{P}(2) + \nicefrac{3}{4} \tilde{P}(1) + \tilde{P}(0) & = 1-c 
  \mathcomma
  \label{eq:probability_equation_system2}\\
  \sum_i^4 \tilde{P}(i)  = 1\mathcomma\label{eq:probability_equation_system3}\\
  0 \leq \tilde{P}(i)  \leq 1\mathcomma\label{eq:probability_equation_system4} 
\end{eqnarray}
where $c=c_\text{Ag}$ and $1-c=c_\text{Pd}$.
This system of equations has, in particular for $c=0.5$, a solution
where the parameter space is spanned by $\tilde{P}(2)$, $\tilde{P}(3)$, and $\tilde{P}(4)$,
under the conditions
\begin{eqnarray}
    \big(2\tilde{P}(2) + 3 \tilde{P}(3)+4\tilde{P}(4) \leq 2\big)
\wedge \nonumber\\
     \bigg\{
      \big[ 
      \big(\tilde{P}(2) + 2\tilde{P}(3)+3\tilde{P}(4)\geq 1\big) 
\wedge 
      \big(\tilde{P}(2)+2\tilde{P}(3)\leq1 \big)
      \big]
\vee \nonumber\\
    \big[\big(\tilde{P}(2) + 2\tilde{P}(3)\geq1 \big) 
    \wedge \big( 2\tilde{P}(2) + 3\tilde{P}(3)\leq2 \big)\big]\bigg\}
\label{eq:probability_c0.50_conditions}
\mathperiod
\end{eqnarray}
The remaining probabilities are then given by
\begin{eqnarray}
    & \tilde{P}(0) = \tilde{P}(2) + 2 \tilde{P}(3) + 3 \tilde{P}(4) -1 \label{eq:probability_c0.50_P0}\mathcomma\\
    & \tilde{P}(1) = 2 - 2\tilde{P}(2) - 3 \tilde{P}(3) - 4 \tilde{P}(4)\label{eq:probability_c0.50_P1}\mathperiod
\end{eqnarray}
If $\tilde{P}(4)$ is zero, a simple solution follows from \eref{eq:probability_c0.50_conditions}
\begin{eqnarray}
  \tilde{P}(0) &= \frac{1-\tilde{P}(2)}{3}\mathcomma \\
  \tilde{P}(1) &= \tilde{P}(4) = 0\mathcomma\\
  \tilde{P}(3) &= -\frac{2(\tilde{P}(2)-1)}{3}\mathcomma
\end{eqnarray}
while $0\leq \tilde{P}(2) \leq 1$ is the only free parameter.
The conditions and probabilities 
for the other concentrations $c=0.25$ and $0.75$ can be found 
following a similar procedure.

\begin{table*}[t]
  \caption{Nearest neighbor SRO parameter and the corresponding configurations used for {fcc} 
    \agpdc{0.25}{0.75} with $N_\text{sub}=4$.
    For $\alpha=0$, all 16 configurations are used and their
    probabilities are given by \eref{eq:probability_alpha0_MS-NL-CPA}.}
  \label{tab:Nsub=4_configurations_c=0.25}
  \lineup
  \begin{indented}
  \item[]
  \begin{center}
  \begin{tabular*}{0.8\textwidth}{l|c|@{\extracolsep{\fill}}*{8}{S} }
    \br
    & & \multicolumn{8}{c}{Probabilities $P(\gamma_i)$ for  $\alpha_1=$}\\
    \multicolumn{1}{c|}{confs.} & \multicolumn{1}{c|}{$c_\text{Ag}$} 
               & \multicolumn{1}{c}{$-\nicefrac{1}{3}$} & -0.201 & -0.141 & -0.095 & \multicolumn{1}{c}{$-\nicefrac{1}{12}$} 
               & -0.04 & 0.099 & \multicolumn{1}{c}{$\quarter$}\\
    \mr
    1 1 1 1 & $1$         & 0    & 0   & 0.05 & 0.05& 0                 & 0.1 & 0.15& $\quarter$  \\
    1 1 1 0 & $\tquarter$ & 0    & 0.05& 0    & 0.05& 0                 & 0   & 0   & 0           \\
    1 1 0 0 & $\half$     & 0    & 0.05& 0.05 & 0   & $\nicefrac{1}{6}$ & 0   & 0.05& 0           \\
    1 0 0 0 & $\quarter$  & 1    & $\tquarter$& 0.7  & 0.65& $\quarter$        & 0.6 & 0.3 & 0           \\
    0 1 0 0 & $\quarter$  & 0    & 0   & 0    & 0   & $\quarter$        & 0   & 0   & 0           \\
    0 0 1 0 & $\quarter$  & 0    & 0   & 0    & 0   & $\quarter$        & 0   & 0   & 0           \\
    0 0 0 1 & $\quarter$  & 0    & 0   & 0    & 0   & $\quarter$        & 0   & 0   & 0           \\
    0 0 0 0 & $0$         & 0    & 0.15& 0.2  & $\quarter$& 0                 & 0.3 & $\half$ & $\tquarter$ \\
    \br
  \end{tabular*}
  \end{center}
  \end{indented}
\end{table*}

\begin{table*}[t]
  \caption{Nearest neighbor SRO parameter and the corresponding configurations used for {fcc} 
    \agpdc{0.5}{0.5} with $N_\text{sub}=4$.
    For $\alpha=0$, all 16 configurations are used and their
    probabilities are given by \eref{eq:probability_alpha0_MS-NL-CPA}.}
  \label{tab:Nsub=4_configurations_c=0.5}
  \lineup
  \begin{indented}
  \item[]
  \begin{center}
  \begin{tabular*}{0.8\textwidth}{l|c|@{\extracolsep{\fill}}*{8}{c} }
    \br
      & & \multicolumn{8}{c}{Probabilities $P(\gamma_i)$ for $\alpha_1 =$}\\
      \multicolumn{1}{c|}{confs.} & \multicolumn{1}{c|}{$c_\text{Ag}$} &  \multicolumn{1}{c}{$-0.270$}
                 & -0.198 &\multicolumn{1}{c}{$-0.140$} & \multicolumn{1}{c}{$-\nicefrac{1}{9}$}  & \multicolumn{1}{c}{$-\nicefrac{1}{12}$} & -0.04 & 0.108 & \multicolumn{1}{c}{$\quarter$}\\
    \mr
    1 1 1 1 & 1          & 0    & 0.1 & 0    & 0                 & 0                 & 0.2 & 0.3 & $\half$ \\%
    1 1 1 0 & $\tquarter$& 0.05 & 0   & 0.15 & 0                 & 0                 & 0   & 0.1 & 0 \\
    1 1 0 1 & $\tquarter$& 0.05 & 0   & 0.15 & 0                 & 0                 & 0   & 0   & 0 \\
    1 1 0 0 & $\half$    & 0.8  & 0.7 & 0.5  & 0                 & $\nicefrac{1}{6}$ & 0.6 & 0.2 & 0 \\
    1 0 1 0 & $\half$    & 0    & 0   & 0    & 0                 & $\nicefrac{1}{6}$ & 0   & 0   & 0 \\
    1 0 0 1 & $\half$    & 0    & 0   & 0    & $\nicefrac{1}{3}$ & $\nicefrac{1}{6}$ & 0   & 0   & 0 \\
    0 1 1 0 & $\half$    & 0    & 0   & 0    & 0                 & $\nicefrac{1}{6}$ & 0   & 0   & 0 \\
    0 1 0 1 & $\half$    & 0    & 0   & 0    & $\nicefrac{1}{3}$ & $\nicefrac{1}{6}$ & 0   & 0   & 0 \\
    0 0 1 1 & $\half$    & 0    & 0   & 0    & $\nicefrac{1}{3}$ & $\nicefrac{1}{6}$ & 0   & 0   & 0 \\
    1 0 0 0 & $\quarter$ & 0.1  & 0.2 & 0.1  & 0                 & 0                 & 0   & 0.1 & 0 \\
    0 0 0 0 & 0          & 0    & 0   & 0.1  & 0                 & 0                 & 0.2 & 0.3 & $\half$ \\%
    \br
  \end{tabular*}
  \end{center}
  \end{indented}
\end{table*}

\section{Used configurations and probabilities}

The configurations and probabilities used to calculate the DOS shown in 
\fref{fig:c25} to \ref{fig:c75} are presented in 
\tref{tab:Nsub=4_configurations_c=0.25}, 
\ref{tab:Nsub=4_configurations_c=0.5}, and
\ref{tab:Nsub=4_configurations_c=0.75}, respectively.
Besides, the 5 representative configurations indicated in 
\tref{tab:Nsub=4_configurations}, we chose also additional 
configurations in order to test the method.

\begin{table}[h]
  \caption{Nearest neighbor SRO parameter and the corresponding configurations used for {fcc}
    \agpdc{0.75}{0.25} with $N_\text{sub}=4$.
    For $\alpha=0$, all 16 configurations are used and their
    probabilities are given by \eref{eq:probability_alpha0_MS-NL-CPA}.}
  \label{tab:Nsub=4_configurations_c=0.75}
  \lineup
  \begin{indented}
  \item[]
  \begin{center}
  \begin{tabular*}{1\columnwidth}{l|c|@{\extracolsep{\fill}}*{6}{c} }
    \br
       & & \multicolumn{6}{c}{Probabilities $P(\gamma_i)$ for $\alpha_1 =$ }\\
    \multicolumn{1}{c|}{confs.} & \multicolumn{1}{c|}{$c_\text{Ag}$} 
               & -0.238 & -0.141 & \multicolumn{1}{c}{$-\nicefrac{1}{12}$}
               & -0.04 & 0.106 & \multicolumn{1}{c}{$\quarter$}\\
    \mr
     1 1 1 1 & 1                 & 01   & 0.2 & 0                 & 0.3 & 0.6 & $\tquarter$ \\%
     1 1 1 0 & $\tquarter$ & 0.85 & 0.7 & $\quarter$ & 0.6 & 0.1 & 0 \\
     1 1 0 1 & $\tquarter$ & 0    & 0   & $\quarter$ & 0   & 0   & 0 \\
     1 0 1 1 & $\tquarter$ & 0    & 0   & $\quarter$ & 0   & 0   & 0 \\ 
     0 1 1 1 & $\tquarter$ & 0    & 0   & $\quarter$ & 0   & 0   & 0 \\ 
     1 1 0 0 & $\half$ & 0    & 0.05& 0                 & 0   & 0.05& 0 \\
     1 0 0 0 & $\quarter$ & 0.05 & 0   & 0                 & 0   & 0.2 & 0 \\                        
     0 0 0 0 & 0                 & 0    & 0.05& 0                 & 0.1 & 0.05& $\quarter$           \\%
     \br
  \end{tabular*}     
  \end{center}
  \end{indented}
\end{table}

\section*{Bibliography}
\bibliography{./journals,./lib}
\bibliographystyle{unsrt}

\end{document}